\documentclass[a4paper]{amsart}
\RequirePackage{amsmath}
\RequirePackage{bm}
\RequirePackage{amssymb}
\RequirePackage{upref}
\RequirePackage{amsthm}
\RequirePackage{enumerate}
\RequirePackage{pb-diagram}
\RequirePackage{amsfonts}
\RequirePackage[mathscr]{eucal}
\RequirePackage{verbatim}
\RequirePackage{xr}
\RequirePackage{graphicx}
\usepackage{calc}
\usepackage{xspace}
\RequirePackage{color}
\RequirePackage{ifthen}

\newcommand{\cf}{cf.\@\xspace}
\newcommand{\resp}{resp.\@\xspace}

\newcommand{\al}{\alpha}
\newcommand{\bet}{\beta}
\newcommand{\ga}{\gamma}
\newcommand{\de}{\delta }

\newcommand{\f}{\varphi}

\newcommand{\lam}{\lambda}
\newcommand{\m}{\mu}

\newcommand{\om}{\omega}

\newcommand{\vt}{\vartheta}

\newcommand{\s}{\sigma}

\newcommand{\D}{\varDelta}
\newcommand{\F}{\varPhi}
\newcommand{\Lam}{\varLambda}
\newcommand{\Om}{\varOmega}

\newcommand{\so}{{\mc S_0}}

\newcommand{\const}{\tup{const}}

\newcommand{\msp[1]}[1]{\mspace{#1mu}}

\newcommand{\R}[1][n+1]{{\protect\mathbb R}^{#1}}

\newcommand{\Ss}[1][n+1]{{\protect\mathbb S}^{#1}}
\newcommand{\Cc}{{\protect\mathbb C}}

\newcommand{\N}{{\protect\mathbb N}}

\newcommand{\Z}{{\protect\mathbb Z}}
\newcommand{\eR}{\stackrel{\lower1ex \hbox{\rule{6.5pt}{0.5pt}}}{\msp[3]\R[]}}
\newcommand{\eN}{\stackrel{\lower1ex \hbox{\rule{6.5pt}{0.5pt}}}{\msp[1]\N}}
\newcommand{\eO}{\stackrel{\lower1ex \hbox{\rule{6pt}{0.5pt}}}{\msc O}}

\newcommand\ra{\rightarrow}

\newcommand\hra{\hookrightarrow}

\newcommand\pde[2]{\frac {\partial#1}{\partial#2}}
\newcommand{\un}{\infty}
\newcommand{\A}{\forall}

\newcommand{\uu}{\cup}
\newcommand{\ii}{\cap}
\newcommand{\uuu}{\bigcup}

\newcommand{\uud}{ \stackrel{\lower 1ex \hbox {.}}{\uu}}
\newcommand{\uuud}[1]{ \stackrel{\lower 1ex \hbox {.}}{\uuu_{#1}}}
\newcommand\su{\subset}

\newcommand\eS{\emptyset}
\newcommand{\sminus}[1][28]{\raise 0.#1ex\hbox{$\scriptstyle\setminus$}}

\newcommand{\abs}[1]{\lvert#1\rvert}

\newcommand{\tbf}{\textbf}
\newcommand{\tit}{\textit}

\newcommand{\tup}{\textup}% text upright

\newcommand{\mc}{\protect\mathcal}
\newcommand{\msc}{\protect\mathscr}

\providecommand{\bysame}{\makebox[3em]{\hrulefill}\thinspace}

\newcommand{\bt}{\begin{thm}}
\newcommand{\bl}{\begin{lem}}
\newcommand{\bc}{\begin{cor}}
\newcommand{\bd}{\begin{definition}}
\newcommand{\bpp}{\begin{prop}}
\newcommand{\br}{\begin{rem}}
\newcommand{\bn}{\begin{note}}
\newcommand{\be}{\begin{ex}}
\newcommand{\bes}{\begin{exs}}
\newcommand{\bb}{\begin{example}}
\newcommand{\bbs}{\begin{examples}}
\newcommand{\ba}{\begin{axiom}}
\newcommand{\bas}{\begin{assumption}}

\newcommand{\et}{\end{thm}}
\newcommand{\el}{\end{lem}}
\newcommand{\ec}{\end{cor}}
\newcommand{\ed}{\end{definition}}
\newcommand{\epp}{\end{prop}}
\newcommand{\er}{\end{rem}}
\newcommand{\en}{\end{note}}
\newcommand{\ee}{\end{ex}}
\newcommand{\ees}{\end{exs}}
\newcommand{\eb}{\end{example}}
\newcommand{\ebs}{\end{examples}}
\newcommand{\ea}{\end{axiom}}
\newcommand{\eas}{\end{assumption}}

\newcommand{\bp}{\begin{proof}}
\newcommand{\ep}{\end{proof}}
\newcommand{\eps}{\renewcommand{\qed}{}\end{proof}}

\newcommand{\bal}{\begin{align}}

\newcommand{\bi}[1][1.]{\begin{enumerate}[\upshape #1]}
\newcommand{\bia}[1][(1)]{\begin{enumerate}[\upshape #1]}
\newcommand{\bin}[1][1]{\begin{enumerate}[\upshape\bfseries #1]}
\newcommand{\bir}[1][(i)]{\begin{enumerate}[\upshape #1]}
\newcommand{\bic}[1][(i)]{\begin{enumerate}[\upshape\hspace{2\cma}#1]}
\newcommand{\bis}[2][1.]{\begin{enumerate}[\upshape\hspace{#2\parindent}#1]}
\newcommand{\ei}{\end{enumerate}}

\newcommand\ndots{\raise 0.47ex \hbox {,}\hskip0.06em\cdots %
     \raise 0.47ex \hbox {,}\hskip0.06em} 

\newcommand{\q}{\quad}
\newcommand{\qq}{\qquad}

\newcommand\nd{\noindent}

\newskip\Csmallskipamount                                                
\Csmallskipamount=\smallskipamount
\newskip\Cmedskipamount
\Cmedskipamount=\medskipamount
\newskip\Cbigskipamount
\Cbigskipamount=\bigskipamount

\newcommand\cvs{\vspace\Csmallskipamount}   
\newcommand\cvm{\vspace\Cmedskipamount}

\newskip\csa
\csa=\smallskipamount

\newskip\cma
\cma=\medskipamount

\newskip\cba
\cba=\bigskipamount

\newdimen\spt
\newcommand\citem{\cvs\advance\itemno by
1{(\romannumeral\the\itemno})\hskip3pt}
\newcommand{\bitem}{\cvm\nd\advance\itemno by
1{\bf\the\itemno}\hspace{\cma}}

\newcount\itemno
\newcommand{\lae}[1]{\label{E:#1}}
\newcommand{\lat}[1]{\label{T:#1}}
\newcommand{\lal}[1]{\label{L:#1}}

\newcommand{\re}[1]{\eqref{E:#1}}

\newcommand{\frl}[1]{Lemma~\ref{L:#1} on page~\tup{\pageref{L:#1}}}

\newcommand{\fre}[1]{\eqref{E:#1} on page~\tup{\pageref{E:#1}}}

\newskip\thmskip
\thmskip=\parindent

\newskip\hsk
\newenvironment{hinw}{\labelsep=0pt\begin{list}{}{\labelsep=0pt\itemindent=0pt\labelwidth=0pt\leftmargin=\parindent\rightmargin=0pt\partopsep=\cba}%
\item\it\nopagebreak\nopagebreak}%
{\end{list}}

\newcommand\bh{\begin{hinw}}
\newcommand{\eh}{\end{hinw}}

\newtheoremstyle{normal}% name
  {\cba}%      Space above, empty = `usual value'
  {\cba}%      Space below
  {}% Body font
  {\thmskip}%Indent amount (empty = no indent, \parindent = para indent)
  {\bfseries}% Thm head font
  {.}%        Punctuation after thm head
  {\hsk}%     Space after thm head: " " = normal interword space;
  {}% Thm head spec

\newtheoremstyle{abschnitt}% name
  {\cba}%      Space above, empty = `usual value'
  {\cba}%      Space below
  {}% Body font
  {\thmskip}% Indent amount (empty = no indent, \parindent = para indent)
  {\bfseries}% Thm head font
  {.}%        Punctuation after thm head
  {\hsk}%     Space after thm head: " " = normal interword space;
  {}% Thm head spec

\newtheoremstyle{italic}% name
  {\cba}%      Space above, empty = `usual value'
  {\cba}%      Space below
  {\itshape}% Body font
  {\thmskip}%  Indent amount (empty = no indent, \parindent = para indent)
  {\bfseries}% Thm head font
  {.}%        Punctuation after thm head
  {\hsk}%     Space after thm head: " " = normal interword space;
  {}% Thm head spec

\newtheoremstyle{aufgaben}% name
  {\cba}%      Space above, empty = `usual value'
  {\cba}%      Space below
  {}% Body font
  {}%         Indent amount (empty = no indent, \parindent = para indent)
  {\normalsize\bfseries}% Thm head font
  {.}%        Punctuation after thm head
  {\hsk}%     Space after thm head: " " = normal interword space;
  {}% Thm head spec

\newtheoremstyle{break}% name
  {\cba}%      Space above, empty = `usual value'
  {\cba}%      Space below
  {\itshape}% Body font
  {}%         Indent amount (empty = no indent, \parindent = para indent)
  {\bfseries}% Thm head font
  {.}%        Punctuation after thm head
  {\newline}% Space after thm head: \newline = linebreak
  {}%         Thm head spec

\swapnumbers
\theoremstyle{italic}
\newtheorem{thm}[subsection]{Theorem}
\newtheorem{lem}[subsection]{Lemma}
\newtheorem{prop}[subsection]{Proposition}
\newtheorem{cor}[subsection]{Corollary}

\theoremstyle{normal}
\newtheorem{rem}[subsection]{Remark}
\newtheorem{definition}[subsection]{Definition}
\newtheorem{example}[subsection]{Example}
\newtheorem{examples}[subsection]{Examples}
\newtheorem{ex}[subsection]{Exercise}
\newtheorem{note}[subsection]{}
\newtheorem{axiom}[subsection]{Axiom}
\newtheorem{assumption}[subsection]{Assumption}

\theoremstyle{aufgaben}
\newtheorem{exs}[subsection]{Exercises}

\swapnumbers

\numberwithin{equation}{section}
\numberwithin{figure}{section}

\newenvironment{textequation}[1][0.8]
{\begin{equation}
\begin{aligned}
\begin{minipage}{#1\linewidth}}
{\end{minipage}
\end{aligned}
\end{equation}
\ignorespacesafterend}

\newcommand{\btext}{\begin{textequation}}
\newcommand{\etext}{\end{textequation}}

\def\hinweis{\@startsection{subsection}{2}%
 \z@{0.7\linespacing\@plus 0.5\linespacing}{0.7\linespacing}%
{\normalfont\itshape\indent}}

\newcounter{hours}\newcounter{minutes}
\newcommand{\printtime}{%
\setcounter{hours}{\time/60}%
\setcounter{minutes}{\time-\value{hours}*60}%
\ifthenelse{\value{minutes}<10}{\thehours :0\theminutes}{\thehours:\theminutes}}

\usepackage[german,english]{babel}
\usepackage{graphicx}
\RequirePackage{amsmath}
\RequirePackage{bm}
\RequirePackage{amssymb}
\RequirePackage{upref}
\RequirePackage{amsthm}
\RequirePackage{enumerate}
\RequirePackage{pb-diagram}
\RequirePackage{amsfonts}
\RequirePackage[mathscr]{eucal}
\RequirePackage{verbatim}
\RequirePackage{xr}
\RequirePackage{graphicx}
\usepackage{calc}
\usepackage{xspace}
\usepackage{dsfont}

\makeatletter
\RequirePackage{color}
\newcommand{\ann}[1]{\renewcommand{\@makefnmark}{\mbox{$^{\color{red}{\@thefnmark}}$}}%
\footnote {#1}}
\makeatother

\RequirePackage{upref}
\RequirePackage{amsthm}
\RequirePackage{enumerate}%\begin{enumerate}[(i)]
\usepackage[mathscr]{eucal}
\usepackage{xr-hyper}

\usepackage{calc}

\newlength{\oddsidemarginlength}
\newlength{\topmarginlength}

\newcounter{numberoflines}
\newcounter{tempcc}
\oddsidemargin=\oddsidemarginlength
\evensidemargin=\oddsidemargin
\topmargin=\topmarginlength
\usepackage[colorlinks=true,linkcolor=blue,citecolor=blue,urlcolor=blue]{hyperref}  

\begin{document}

\flushbottom

%\larger[1]
%\frontmatter

\title{The quantization of a Kerr-AdS black hole}

% author one information
\author{Claus Gerhardt}
\address{Ruprecht-Karls-Universit\"at, Institut f\"ur Angewandte Mathematik,
Im Neuenheimer Feld 205, 69120 Heidelberg, Germany}
%\curraddr{}
\email{\href{mailto:gerhardt@math.uni-heidelberg.de}{gerhardt@math.uni-heidelberg.de}}
\urladdr{\href{http://www.math.uni-heidelberg.de/studinfo/gerhardt/}{http://www.math.uni-heidelberg.de/studinfo/gerhardt/}}
%\thanks{This work was supported by the DFG}

% author two information
%\author{}
%\address{}
%\curraddr{}
%\email{}
%\thanks{}
%
\date{\today}
%
% at present the "communicated by" line appears only in ERA and PROC
%\commby{}

%\dedicatory{}

\begin{abstract} 
We apply our model of quantum gravity to a Kerr-AdS spacetime of dimension $2 m+1$, $\ge2$, where all rotational parameters are equal, resulting in a wave equation in a quantum spacetime  which has a sequence of solutions that can be expressed as a product of stationary and temporal eigenfunctions. The stationary eigenfunctions can be interpreted as radiation and the temporal as gravitational waves. The event horizon corresponds in the quantum model to a Cauchy hypersurface that can be crossed by causal curves in both directions such that the information paradox does not occur. We also prove that the Kerr-AdS spacetime can be maximally extended by replacing in a generalized Boyer-Lindquist coordinate system the $r$ variable by $\rho=r^2$ such that the extended spacetime has a timelike curvature singularity in $\rho=-a^2$.
\end{abstract}

\maketitle

\tableofcontents

\setcounter{section}{0}
\section{Introduction}
In general relativity the Cauchy development of a Cauchy hypersurface $\so$ is governed by the Einstein equations, where of course the second fundamental form of $\so$ has also to be specified.

 In the model of quantum gravity  we developed in a series of papers \cite{cg:qgravity,cg:uqtheory,cg:uqtheory2,cg:qgravity2, cg:uf2, cg:uf3,cg:uf4} we pick a Cauchy hypersurface, which is then only considered to be a complete Riemannian manifold $(\so,g_{ij})$ of dimension $n\ge 3$, and define its quantum development to be special  solutions of the wave equation
\begin{equation}\lae{1.1}
\frac1{32}\frac{n^2}{n-1}\Ddot u
-(n-1) t^{2-\frac4n}\D u-\frac {n}2 t^{2-\frac4n}Ru+nt^2\Lam u=0
\end{equation}
defined in the spacetime
\begin{equation}
Q=\so\times (0,\un).
\end{equation}
The Laplacian is the Laplacian with respect to $g_{ij}$, $R$ is the scalar curvature of the metric, $0<t$ is the time coordinate defined by the derivation process of the equation and $\Lam<0$ a cosmological constant. If other physical fields had been present in the Einstein equations then the  equation would contain further lower order terms, \cf \cite{cg:uf2}, but a negative cosmological constant would always have to be present even if the Einstein equations would only describe the vacuum.

Using separation of variables we proved that there is a complete sequence of eigenfunctions $v_j$ (or a complete set of eigendistributions) of a stationary eigenvalue problem and a complete sequence of eigenfunctions $w_i$ of a temporal  implicit eigenvalue problem, where $\Lam$ plays the role of an eigenvalue, such that the functions
\begin{equation}
u(t,x)=w_i(t)v_j(x)
\end{equation}
are solutions of the wave equation, \cf \cite[Section 6]{cg:qgravity2} and \cite{cg:uf2,cg:uf3}.

In a recent  paper \cite{cg:qbh} we applied this model to quantize a Schwarzschild-AdS spacetime and we shall prove in the present paper that similar arguments can also be used to quantize a Kerr-AdS spacetime with a rotating black hole. 

We consider an odd-dimensional Kerr-AdS spacetime $N$, $\dim N=2m+1$, $m\ge 2$, where all rotational parameters are equal
\begin{equation}
a_i=a\qq\A\, 1\le i\le m,
\end{equation}
and where we also set
\begin{equation}
n=2m.
\end{equation}
Replacing the $r$ coordinate in a generalized Boyer-Lindquist coordinate system by
\begin{equation}
\rho=r^2
\end{equation}
we shall prove that in the new coordinate system the metric is smooth in the interval
\begin{equation}
-a^2<\rho<\un
\end{equation}
and that the extended spacetime $N$ has a timelike curvature singularity in $\rho=-a^2$, \cf \frl{2.6}. 

For the quantization we first assume that there is a non-empty interior black hole region  $B$ which is bounded by two horizons
\begin{equation}
B=\{r_1<r<r_2\}
\end{equation}
where the outer horizon is the event horizon. Picking a Cauchy hypersurface in $B$ of the form
\begin{equation}
\{r=\const\},
\end{equation}
 we shall prove  that the induced metric of the Cauchy hypersurface can be expressed in the form
\begin{equation}\lae{1.5}
ds^2=d\tau^2+\s_{ij}dx^idx^j,
\end{equation}
where
\begin{equation}
-\un<\tau<\un,
\end{equation}
$r=\const$ and $\s_{ij}$ is a smooth Riemannian metric on $\Ss[2m-1]$ depending on $r,a$ and the cosmological constant $\Lam<0$. The metric in \re{1.5} is free of any coordinate singularity, hence we can let $r$ tend to $r_2$ such that the Cauchy hypersurfaces converge to a Riemannian manifold $\so$ which represents the event horizon at least topologically. Furthermore, the Laplacian of the metric in \re{1.5} comprises a harmonic oscillator with respect to $\tau$ which enables us to write the stationary eigenfunctions $v_j$ in the form
\begin{equation}
v_j(\tau,x)=\zeta(\tau)\f_j(x),
\end{equation}
where $\f_j$ is an eigenfunction of the elliptic operator
\begin{equation}\lae{1.12}
-(n-1)\tilde\D -\frac n2 R,
\end{equation}
where
\begin{equation}
\tilde \D= \D_M,
\end{equation}
 and $\zeta$ an eigenfunction of the harmonic oscillator the frequency of which are still to be determined. 

Due to the presence of the harmonic oscillator we can now consider an \tit{explicit} temporal eigenvalue problem, i.e., we consider the eigenvalue problem
\begin{equation}\lae{1.9}
-\frac1{32} \frac{n^2}{n-1}\Ddot w+n\abs\Lam t^2w=\lam t^{2-\frac4n}w
\end{equation}
with a fixed $\Lam<0$, where we choose $\Lam$ to be the cosmological constant of the Kerr-AdS spacetime.

The eigenvalue problem \re{1.9} has a complete sequence $(w_i,\lam_i)$ of eigenfunctions with finite energies $\lam_i$ such that
\begin{equation}
0<\lam_0<\lam_1<\cdots
\end{equation}
and by choosing the frequencies of $\zeta$ appropriately we can arrange that the stationary eigenvalues $\mu_j$ of $v_j$ agree with the temporal eigenvalues $\lam_i$. If this is the case then the eigenfunctions
\begin{equation}
u=w_iv_j
\end{equation}
will be a solution of the wave equation. More precisely we proved: 
\bt
Let $(\f_j,\tilde\mu_j)$ \resp $(w_i,\lam_i)$ be eigenfunctions of the elliptic operator in \re{1.12} 
 \resp the temporal eigenfunctions and, for a given index $j$, let $\lam_{i_0}$ be the smallest eigenvalue of the $(\lam_i)$ with the property
\begin{equation}
\lam_{i_0}\ge \tilde\mu_j,
\end{equation}
then, for any $i\ge i_0$, there exists 
\begin{equation}
\om=\om_{ij}\ge 0
\end{equation}
and corresponding $\zeta_{ij}$ satisfying
\begin{equation}
-\Ddot\zeta_{ij}=\om_{ij}^2
\end{equation}
such that
\begin{equation}
\lam_i=\mu_{ij}=\om_{ij}^2+\tilde\mu_j\q\A\, i\ge i_0.
\end{equation}
The functions
\begin{equation}
u_{ij}=w_i\zeta_{ij}\f_j
\end{equation}
are then solutions of the wave equation with bounded energies satisfying
\begin{equation}
\lim_{t\ra 0}u_{ij}(t)=\lim_{t\ra\un}u_{ij}(t)=0
\end{equation}
and
\begin{equation}
u_{ij}\in C^\un(\R[*]_+\times\so)\ii C^{2,\al}(\bar{\R[]}^*_+\times \so)
\end{equation}
for some
\begin{equation}
\frac23\le\al<1.
\end{equation}

Moreover, we have
\begin{equation}
\om_{ij}>0\qq\A\, i>i_0.
\end{equation}
If
\begin{equation}
\lam_{i_0}=\tilde\mu_j,
\end{equation}
then we define
\begin{equation}
\zeta_{i_0j}\equiv 1.
\end{equation}
\et
\br
(i) The event horizon corresponds to the Cauchy hypersurface $\{t=1\}$ in $Q$ and the open set 
\begin{equation}
\{-a^2<\rho<\rho_2\}
\end{equation}
in $N$, where
\begin{equation}
\rho_2=r_2^2,
\end{equation}
  to the region
\begin{equation}
\so\times (0,1),
\end{equation}
while the part 
\begin{equation}
\{\rho_2<\rho<\un\}
\end{equation}
 is represented by
\begin{equation}
\so\times (1,\un).
\end{equation}
The \tit{timelike} black hole singularity corresponds to $\{t=0\}$ which is  a \tit{spacelike} curvature singularity in the quantum spacetime provided we equip $Q$ with a metric such that the hyperbolic operator is normally hyperbolic, \cf \cite[Lemma 6.2]{cg:uf2}. Moreover, in the quantum spacetime the Cauchy hypersurface $\so$ can be crossed by causal curves in both directions, i.e., the information paradox does not occur.

\cvm
(ii) The stationary eigenfunctions can be looked at as being radiation because they comprise the harmonic oscillator, while we consider the temporal eigenfunctions to be gravitational waves.
\er
As it is well-known the Schwarzschild black hole or more specifically the extended Schwarzschild space has already been analyzed by Hawking \cite{hawking:bh} and Hartle and Hawking \cite{hartle-hawking}, see also the book by Wald \cite{wald:qft}, using quantum field theory, but not quantum gravity, to prove that the black hole emits radiation.

The metric describing a rotating black hole in a four-dimensional vacuum spacetime was first discovered by Kerr \cite{kerr:metric}. Carter \cite{carter:kerr} generalized the Kerr solution by describing a rotating black hole in a four-dimensional de Sitter or anti-de Sitter background. Higher dimensional solutions for a rotating black hole were given by Myers and Perry \cite{myers:perry} in even-dimensional Ricci flat spacetimes and by Hawking, Hunter and Taylor \cite{hawking:kerr} in five-dimensional spacetimes satisfying the Einstein equations with cosmological constant.

A general solution in all dimension was given in \cite{gibbons:kerr-ads} by Gibbons et al. and we shall use their metric in odd dimensions, with all rotational parameters supposed to be equal, to define our spacetime $N$, though we shall maximally extend it.

\bn{\tbf{Notations.}} We apply the summation convention and label coordinates with contravariant indices, e.g., $\m^i$. However, for better readability we  shall usually write 
\begin{equation}
\m_i^2
\end{equation}
instead of
\begin{equation}
(\m^i)^2.
\end{equation}
\en

\section{Preparations}

We consider odd-dimensional Kerr-AdS spacetimes $N$, $\dim N=2m+1$, $m\ge 2$, assuming that all rotational parameters are equal
\begin{equation}
a_i=a\not=0,\qq\A\,1\le i\le m.
\end{equation}
The Kerr-Schild form of the metric can then be expressed as
\begin{equation}\lae{2.2}
\begin{aligned}
d\bar s^2&= -\frac{1+l^2r^2}{1-a^2l^2}dt^2+\frac{r^2dr^2}{(1+l^2r^2)(r^2+a^2)}\\
&\q +\frac{r^2+a^2}{1-a^2l^2}\sum_{i=1}^m(d\m_i^2+\m_i^2d\f_i^2)\\
&\q+\frac{2m_0}U\big(\frac1{1-a^2l^2}(dt-a\sum_{i=1}^m\m_i^2d\f^i)+\frac{r^2dr}{(1+l^2r^2)(r^2+a^2)}\big)^2,
\end{aligned}
\end{equation}
where
\begin{equation}
l^2=-\frac1{m(2m-1)}\Lam
\end{equation}
and $\Lam<0$ is the cosmological constant such that the Einstein equations
\begin{equation}
G_{\al\bet}+\Lam\bar g_{\al\bet}=0
\end{equation}
are satisfied in $N$, $m_0$ is the mass of the black hole,
\begin{equation}
U=(r^2+a^2)^{m-1},
\end{equation}
\begin{equation}
\sum_{i=1}^m(d\m_i^2+\m_i^2d\f_i^2)
\end{equation}
is the standard metric of $\Ss[2m-1]$, where the $\f^i$ are the azimuthal coordinates, the values of which have to be identified modulo $2\pi$, and the $\m^i$ are the latitudinal coordinates subject to the side-condition
\begin{equation}\lae{2.7}
\sum_{i=1}^m\m_i^2=1.
\end{equation}
The $\m^i$ also satisfy
\begin{equation}
0\le\m^i\le 1\qq\A\, 1\le i\le m.
\end{equation}
The coordinates $(t,r)$ are defined in 
\begin{equation}
-\un<t<\un
\end{equation}
and
\begin{equation}
0<r<\un
\end{equation}
respectively, \cf \cite[Section 2 and Appendix B]{gibbons:kerr-ads}.

The horizons are hypersurfaces $\{r=\const\}$, where $\rho=r^2$ satisfies the equation
\begin{equation}\lae{2.11}
(1+l^2\rho)(\rho+a^2)^m-2m_0\rho=0.
\end{equation}
Let
\begin{equation}\lae{2.12}
\F=\F(\rho)
\end{equation}
be the polynomial on the left-hand side of \re{2.11}, then $\F$ is strictly convex in $\R[]_+$ and we have
\begin{equation}
\F(0)>0
\end{equation} 
and
\begin{equation}
\lim_{\rho\ra\un}\F(\rho)=\un,
\end{equation}
from which we deduce that the equation \re{2.11} is satisfied if and only if
\begin{equation}
\inf_{\R[]_+}\F\le0,
\end{equation}
and in case
\begin{equation}
\inf_{\R[]_+}\F<0
\end{equation}
we have exactly two solutions otherwise only one. If there are two solutions $r_i$, $i=1,2$, such that
\begin{equation}
0<r_1<r_2,
\end{equation}
then the outer horizon is called \tit{event horizon} and the black hole has an interior region
\begin{equation}
B=\{r_1<r<r_2\}
\end{equation}
in which the variable $r$ is a time coordinate. If there is only one solution $r_0$, then $B$ is empty and the black hole is called \tit{extremal}.

We shall first quantize a black hole with $B\not=\eS$; the quantization of an extremal black hole is then achieved by approximation.

Thus, let us consider a non-extremal black hole and let $S\su B$ be a spacelike coordinate slice
\begin{equation}
S=S(r)=\{r=\const\},
\end{equation}
where $r$ also denotes the constant value.

In view of  \re{2.2}, the induced metric can be expressed as 
\begin{equation}\lae{2.20}
\begin{aligned}
ds_S^2&=\big(\frac{2m_0}U\frac1{(1-a^2l^2)^2}-\frac{1+l^2 r^2}{1-a^2 l^2}\big)dt^2-\frac{2m_0}U\frac{2a}{(1-a^2 l^2)^2}\mu_i^2dtd\f^i\\
&\q+\big(\frac{2m_0}U\frac{a^2}{(1-a^2 l^2)^2}\m_i^2\m_j^2+\frac{r^2+a^2}{1-a^2 l^2}\m_i^2\de_{ij}\big)d\f^i d\f^j\\
&\q+\frac{r^2+a^2}{1-a^2l^2}\sum_{i=1}^md\m_i^2,
\end{aligned}
\end{equation}
from which we deduce
\begin{equation}\lae{2.21}
g_{tt}=\frac{2m_0}U\frac1{(1-a^2l^2)^2}-\frac{1+l^2 r^2}{1-a^2 l^2},
\end{equation}
\begin{equation}\lae{2.22}
g_{t\f^i}=g_{\f^it}=-\frac{2m_0}U\frac{a}{(1-a^2 l^2)^2}\mu_i^2,
\end{equation}
and
\begin{equation}\lae{2.23}
g_{\f^i\f^j}=\frac{2m_0}U\frac{a^2}{(1-a^2 l^2)^2}\m_i^2\m_j^2+\frac{r^2+a^2}{1-a^2 l^2}\m_i^2\de_{ij}.
\end{equation}
The other components of the metric are either $0$ or are represented by the line element
\begin{equation}\lae{2.24} 
\frac{r^2+a^2}{1-a^2l^2}\sum_{i=1}^md\m_i^2,
\end{equation}
note the constraint \re{2.7}.

To eliminate the $g_{t\f^i}$ we shall introduce new coordinates. First, let us make the simple change by defining $t'$ through
\begin{equation}
ct'=t,
\end{equation}
where $c\not=0$ is a constant which will be specified later,  and dropping the prime in the sequel, resulting in a replacement of the components in \re{2.21} and \re{2.22} by 
\begin{equation}\lae{2.26}
g_{tt}=c^2\big(\frac{2m_0}U\frac1{(1-a^2l^2)^2}-\frac{1+l^2 r^2}{1-a^2 l^2}\big)
\end{equation}
respectively,
\begin{equation}\lae{2.27}
g_{t\f^i}=g_{\f^it}=-c\frac{2m_0}U\frac{a}{(1-a^2 l^2)^2}\mu_i^2.
\end{equation}

Next, we define new coordinates $(\tilde t,\tilde \f^i)$ by
\begin{equation}
\al\tilde t=t
\end{equation}
and
\begin{equation}
\tilde\f^i=\f^i-a\ga t,
\end{equation}
where $\al, \ga$ are non-vanishing constants to specified later, such that
\begin{equation}
\f^i=\tilde\f^i+a\al\ga \tilde t.
\end{equation}
In the new coordinates the only interesting new components are
\begin{equation}
\begin{aligned}
g_{\tilde t\tilde t}&=g_{tt}\pde t{\tilde t}\pde t{\tilde t}+2g_{t\f^i}\pde t{\tilde t}\pde{\f^i}{\tilde t}+g_{\f^i\f^j}\pde{\f^i}{\tilde t}\pde{\f^j}{\tilde t}\\
&=\al^2\big(g_{tt}+2a\ga \sum_{i}g_{t\f^i}+a^2\ga^2\sum_{i,j}g_{\f^i\f^j}\big)
\end{aligned}
\end{equation}
and
\begin{equation}
\begin{aligned}
g_{\tilde t\tilde\f^i}&=g_{t\f^j}\pde t{\tilde t}\pde{\f^j}{\tilde\f^i}+g_{\f^k\f^l}\pde{\f^k}{\tilde t}\pde{\f^l}{\tilde\f^i}\\
&=\al\big(g_{t\f^i}+a\ga \sum_kg_{\f^k\f^i}\big).
\end{aligned}
\end{equation}
We therefore deduce, in view of \re{2.23}, \re{2.26} and \re{2.27}, 
\begin{equation}\lae{2.31}
g_{\tilde t\tilde t}=\al^2\big(\frac{2m_0}U\frac1{(1-a^2l^2)^2}(c-a^2\ga)^2+\frac{r^2+a^2}{1-a^2 l^2}a^2\ga^2-\frac{1+l^2 r^2}{1-a^2 l^2}c^2\big)
\end{equation}
and
\begin{equation}
g_{\tilde t\tilde \f^i}=\al\big(\frac{2m_0}U\frac a{(1-a^2 l^2)^2}(a^2 \ga-c)+\frac{r^2+a^2}{1-a^2 l^2}a\ga\big)\m_i^2.
\end{equation}
Choosing now
\begin{equation}\lae{2.35}
c=\big(a^2+\frac U{2m_0}(r^2+a^2)(1-a^2 l^2)\big)\ga
\end{equation}
we conclude
\begin{equation}
g_{\tilde t\tilde\f^i}=0.
\end{equation}
Combining then \re{2.35} and \re{2.31} by setting $\ga=1$ we obtain
\begin{equation}\lae{2.37}
\begin{aligned}
g_{\tilde t\tilde t}&=\al^2\big(\frac U{2m_0}(r^2+a^2)^2+\frac {r^2+a^2}{1-a^2 l^2} a^2\\
&\q-\frac{1+l^2 r^2}{1-a^2 l^2}(a^2+\frac U{2m_0}(r^2+a^2)(1-a^2 l^2))^2\big).
\end{aligned}
\end{equation}
Define
\begin{equation}
\bet=\frac U{2m_0}(r^2+a^2)-\frac{r^2}{1+l^2 r^2},
\end{equation}
then
\begin{equation}\lae{2.39}
\bet<0\qq\text{in }B,
\end{equation}
since the function $\F$ in \re{2.12} is negative in $B$. Writing
\begin{equation}
\begin{aligned}
a^2+\frac U{2m_0}(r^2+a^2)(1-a^2 l^2)&=a^2+\frac{r^2}{1+l^2 r^2}(1-a^2 l^2)\\
&\q+\bet (1-a^2 l^2)\\
&=\frac{r^2+a^2}{1+l^2 r^2}+\bet (1-a^2 l^2),
\end{aligned}
\end{equation}
we infer
\begin{equation}\lae{2.41}
g_{\tilde t\tilde t}=\al^2(-\bet(r^2+a^2)-\bet^2(1+l^2 r^2)(1-a^2 l^2)).
\end{equation}
The term in the brackets vanishes on the event horizon and is strictly positive in $B$, in view of \re{2.39} and the identity 
\begin{equation}
\begin{aligned}
&\msp[9]\q(r^2+a^2)+\bet(1+l^2 r^2)(1-a^2 l^2)\\
&= (r^2+a^2)+\frac{(r^2+a^2)^m}{2m_0}(1+l^2 r^2)(1-a^2 l^2)-r^2(1-a^2 l^2)\\
&=a^2(1+l^2)+\frac{(r^2+a^2)^m}{2m_0}(1+l^2 r^2)(1-a^2 l^2)>0.
\end{aligned}
\end{equation}
Hence, for any $r$ satisfying
\begin{equation}
r_1<r<r_2
\end{equation}
we can choose $\al>0$ such that
\begin{equation}
g_{\tilde t\tilde t}=1.
\end{equation}
Writing $(\tau,\f^i)$ instead of $(\tilde t,\tilde\f^i)$ we can then state
\bl
For any hypersurface
\begin{equation}\lae{2.45}
S=S(r)\su B
\end{equation}
the induced metric can be expressed in the form
\begin{equation}\lae{2.46}
\begin{aligned}
ds_S^2&=d\tau^2+\big(\frac{2m_0}U\frac{a^2}{(1-a^2 l^2)^2}\m_i^2\m_j^2+\frac{r^2+a^2}{1-a^2 l^2}\m_i^2\de_{ij}\big)d\f^i d\f^j\\
&\q+\frac{r^2+a^2}{1-a^2l^2}\sum_{i=1}^md\m_i^2\\
&\equiv d\tau^2+\s_{ij}dx^idx^j,
\end{aligned}
\end{equation}
where 
\begin{equation}
\s_{ij}=\s_{ij}(r,a,l)
\end{equation}
is a smooth Riemannian metric on $\Ss[2m-1]$ and $\tau$ ranges in $\R[]$, while in case
\begin{equation}\lae{2.48}
S=S(r)\su N\sminus \bar B,
\end{equation}
the induced metric is  Lorentzian of the form
\begin{equation}\lae{2.49}
\begin{aligned}
ds_S^2&=-d\tau^2+\big(\frac{2m_0}U\frac{a^2}{(1-a^2 l^2)^2}\m_i^2\m_j^2+\frac{r^2+a^2}{1-a^2 l^2}\m_i^2\de_{ij}\big)d\f^i d\f^j\\
&\q+\frac{r^2+a^2}{1-a^2l^2}\sum_{i=1}^md\m_i^2\\
&\equiv -d\tau^2+\s_{ij}dx^idx^j.
\end{aligned}
\end{equation}
If $r<r_2$ tends to $r_2$, then the hypersurfaces S(r) converge topologically to the event horizon and the induced metrics to the Riemannian metric
\begin{equation}\lae{2.50}
\begin{aligned}
ds_S^2&=d\tau^2+\frac{r^2+a^2}{1-a^2 l^2}\big(\de_{ij}d\m^id\m^j+\m_i^2\de_{ij}d\f^i d\f^j\big)\\
&\q+a^2\frac{(1+l^2 r^2) (r^2+a^2)}{r^2 (1-a^2 l^2)^2}\m_i^2\m_j^2d\f^id\f^j.
\end{aligned}
\end{equation}
\el
\bp
We only have to prove the case \re{2.48}. However, the proof of this case is identical to the proof when \re{2.45} is valid by observing that then the term $\bet$ in \re{2.41} is strictly positive.
\ep

\section{The quantization}
We are now in a position to argue very similar as in our former paper \cite[Section 2]{cg:qbh}. For the convenience of the reader we shall repeat some of the arguments so that the results can be understood directly without having to look up the details in the reference.

The interior of the black hole is a globally hyperbolic spacetime and the slices $S(r)$ with 
\begin{equation}
r_1<r<r_2
\end{equation}
 are Cauchy hypersurfaces. Let $r$ tend to $r_2$ and let $\so$ be the resulting limit Riemannian manifold, i.e., topologically it is the event horizon but equipped with the metric in \re{2.50} which we shall write in the form
 \begin{equation}\lae{2.52}
ds^2=d\tau^2+\s_{ij}dx^idx^j
\end{equation}
as in \fre{2.46}. By a slight abuse of language we shall also call $\so$ to be a Cauchy hypersurface though it is only the geometric limit of Cauchy hypersurfaces. However, $\so$ is a genuine Cauchy hypersurface in the quantum model which is defined by the equation \fre{1.1}.

Let us now look at the stationary eigenvalue equation, where we recall that $n=2m$,
\begin{equation}\lae{2.2.11}
-(n-1)\D v-\frac n2Rv=\mu v
\end{equation}
in $\so$, where
\begin{equation}
-(n-1)\D v=-(n-1)\Ddot v -(n-1) \tilde \D v
\end{equation}
and $\tilde\D$ is the Laplacian in the Riemannian manifold
\begin{equation}
M=(\Ss[n-1],\s_{ij});
\end{equation}
moreover the scalar curvature $R$ is also the scalar curvature with respect to $\s_{ij}$ in view of \re{2.52}.
Using separation of variables let us write
\begin{equation}
v(\tau,x)=\zeta(\tau)\f(x)
\end{equation}
to conclude that the left-hand side of \re{2.2.11} can be expressed in the form
\begin{equation}
-(n-1)\Ddot\zeta\f+\zeta\{-(n-1)\tilde\D\f-\frac n2 R\f\}.
\end{equation}
Hence, the eigenvalue problem \re{2.2.11} can be solved by setting
\begin{equation}
v=\zeta\f_j,
\end{equation}
where $\f_j$, $j\in\N$, is an eigenfunction of the elliptic operator
\begin{equation}\lae{2.59}
-(n-1)\tilde\D-\frac n2 R
\end{equation}
such that
\begin{equation}
-(n-1)\tilde\D\f_j-\frac n2 R\f_j=\tilde\mu_j\f_j,
\end{equation}
\begin{equation}
\tilde\mu_0<\tilde\mu_1\le \tilde\mu_2\le \cdots
\end{equation}
and $\zeta$ is an eigenfunction of the harmonic oscillator. The eigenvalue of the harmonic oscillator can be arbitrarily positive or zero. We define it at the moment as
\begin{equation}
\om^2
\end{equation}
where $\om\ge 0$ will be determined later. For $\om>0$ we shall consider the real eigenfunction
\begin{equation}\lae{2.2.23}
\zeta=\sin \om\tau
\end{equation}
which represents the ground state in the interval 
\begin{equation}
I_0=(0,\frac\pi{\om})
\end{equation}
with vanishing boundary values. $\zeta$ is a solution of the variational problem
\begin{equation}
\frac{\int_{I_0}\abs {\dot\vartheta}^2}{\int_{I_0}\abs\vartheta^2}\ra\min\q\A\, 0\not=\vt\in H^{1,2}_0(I_0)
\end{equation}
in the Sobolev space $H^{1,2}_0(I_0)$.

Multiplying $\zeta$ by a constant we may assume
\begin{equation}
\int_{I_0}\abs\zeta^2=1.
\end{equation}
Obviously,
\begin{equation}
\so=\R[]\times M
\end{equation}
and though $\zeta$ is defined in $\R[]$ and is even an eigenfunction it has infinite norm in $L^2(\R[])$. However, when we consider a finite disjoint union of $N$ open intervals $I_j$
\begin{equation}
\Om=\uuu_{j=1}^NI_j,
\end{equation}
where
\begin{equation}
I_j=(k_j\frac\pi{\om},(k_j+1)\frac\pi{\om}),\q k_j\in\Z,
\end{equation}
then
\begin{equation}
\zeta_N=N^{-\frac12}\zeta
\end{equation}
is a unit eigenfunction in $\Om$ with vanishing boundary values having the same energy as $\zeta$ in $I_0$. Hence, it suffices to consider $\zeta$ only in $I_0$ since this configuration can immediately be generalized to arbitrary large bounded open intervals
\begin{equation}
\Om\su\R[].
\end{equation}
We then can state:
\bl
There exists a complete sequence of unit eigenfunctions of the operator in \re{2.59} with eigenvalues $\tilde\mu_j$ such that the functions
\begin{equation}
v_j=\zeta \f_j,
\end{equation}
where $\zeta$ is a constant multiple of the function in \re{2.2.23} with unit $L^2(I_0)$ norm, are solutions of the eigenvalue problem \re{2.2.11} with eigenvalue 
\begin{equation}\lae{2.2.33}
\mu_j=(n-1)\om^2+\tilde\mu_j.
\end{equation}
The eigenfunctions $v_j$ form an orthogonal basis for $L^2(I_0\times M,\Cc)$. 
\el

To solve the wave equation \fre{1.1} let us first consider the following eigenvalue problem
\begin{equation}\lae{2.2.36}
-\frac1{32} \frac{n^2}{n-1}\Ddot w+n\abs\Lam t^2w=\lam t^{2-\frac4n}w
\end{equation}
in the Sobolev space
\begin{equation}
H^{1,2}_0(\R[*]_+).
\end{equation}
Here, 
\begin{equation}
\Lam<0
\end{equation}
can in principle be an arbitrary negative parameter but in the case of a Kerr-AdS black hole it seems reasonable to choose the cosmological constant of the Kerr-AdS spacetime. 

The eigenvalue problem \re{2.2.36} can be solved by considering the generalized eigenvalue problem for the bilinear forms
\begin{equation}
B(w,\tilde w)=\int_{\R[*]_+}\{\frac 1{32}\frac{n^2}{n-1}\bar w'\tilde w'+n\abs\Lam t^2\bar w\tilde w\}
\end{equation}
and
\begin{equation}
K(w,\tilde w)=\int_{\R[*]_+}t^{2-\frac4n}\bar w\tilde w
\end{equation}
in the Sobolev space $\mc H$ which is the completion of
\begin{equation}
C^\un_c(\R[*]_+,\Cc)
\end{equation}
in the norm defined by the first bilinear form.

We then look at the generalized eigenvalue problem
\begin{equation}\lae{2.2.43}
B(w,\f)=\lam K(w,\f)\q\A\,\f\in\mc H
\end{equation}
which is equivalent to \re{2.2.36}.
\bt\lat{2.2.2}
The eigenvalue problem \re{2.2.43} has countably many solutions $(w_i,\lam_i)$ such that
\begin{equation}\lae{2.2.44}
0<\lam_0<\lam_1<\lam_2<\cdots,
\end{equation}
\begin{equation}
\lim\lam_i=\un,
\end{equation}
and
\begin{equation}
K(w_i,w_j)=\de_{ij}.
\end{equation}
The $w_i$ are complete in $\mc H$ as well as in $L^2(\R[*]_+)$.
\et
\bp
The quadratic form $K$ is compact with respect to the quadratic form $B$ as one can easily prove, \cf \cite[Lemma 6.8]{cg:qfriedman}, and hence a proof of the result, except for the strict inequalities in \re{2.2.44}, can be found in \cite[Theorem 1.6.3, p. 37]{cg:pdeII}. Each eigenvalue has multiplicity one since we have a linear ODE of order two and all solutions satisfy the boundary condition 
\begin{equation}\lae{6.45}
 w_i(0)=0.
\end{equation}
The kernel is two-dimensional and the condition \re{6.45} defines a one-dimen\-sional subspace. Note, that we considered only real valued solutions to apply this argument. 
\ep
We are now ready to define the solutions of the wave equation \re{1.1}. 
\bt
Let $(\f_j,\tilde\mu_j)$ \resp $(w_i,\lam_i)$ be eigenfunctions of the elliptic operator in \re{2.59}
 \resp the temporal eigenfunctions and let $\lam_{i_0}$ be the smallest eigenvalue of the $(\lam_i)$ with the property
\begin{equation}
\lam_{i_0}\ge \tilde\mu_j,
\end{equation}
then, for any $i\ge i_0$, there exists 
\begin{equation}
\om=\om_{ij}\ge 0
\end{equation}
and corresponding $\zeta_{ij}$ satisfying
\begin{equation}
-\Ddot\zeta_{ij}=\om_{ij}^2\zeta_{ij}
\end{equation}
such that
\begin{equation}
\lam_i=\mu_{ij}=(n-1)\om_{ij}^2+\tilde\mu_j\q\A\, i\ge i_0.
\end{equation}
The functions
\begin{equation}
u_{ij}=w_i\zeta_{ij}\f_j
\end{equation}
are then solutions of the wave equation with bounded energies satisfying
\begin{equation}
\lim_{t\ra 0}w_{ij}(t)=\lim_{t\ra\un}w_{ij}(t)=0
\end{equation}
and
\begin{equation}
w_{ij}\in C^\un(\R[*]_+\times\so)\ii C^{2,\al}(\bar{\R[]}^*_+\times \so)
\end{equation}
for some
\begin{equation}
\frac23\le\al<1.
\end{equation}

Moreover, we have
\begin{equation}
\om_{ij}>0\qq\A\, i>i_0.
\end{equation}
If
\begin{equation}
\lam_{i_0}=\tilde\mu_j,
\end{equation}
then we define
\begin{equation}
\zeta_{i_0j}\equiv 1.
\end{equation}
\et
\bp
The proof is obvious.
\ep
\br
(i) By construction the temporal and spatial energies of the solutions of the wave equation have to be equal.

\cvm
(ii) The stationary solutions comprising a harmonic oscillator can be looked at a being radiation while we consider the temporal solutions to be gravitational waves. 

\cvm
(iii) If one wants to replace the bounded Interval $I_0$ by $\R[]$ then the eigenfunctions $\zeta_{ij}$ have to be replaced by eigendistributions. An appropriate choice would be
\begin{equation}
\zeta_{ij}=e^{i\om_{ij}\tau};
\end{equation}
also see \cite{cg:uf3} for a more general setting.
\er

The hyperbolic operator defined by the wave equation \fre{1.1} can be defined in the spacetime
\begin{equation}
Q=\so\times\R[*]_+
\end{equation}
which can be equipped with the Lorentzian metrics
\begin{equation}\lae{6.17}
d\bar s^2=-\frac{32(n-1)}{n^2}dt^2+g_{ij}dx^idx^j
\end{equation}
as well as with the metric
\begin{equation}\lae{6.18}
d\tilde s^2=-\frac{32(n-1)}{n^2}dt^2+\frac1{n-1}t^{\frac4n-2}g_{ij}dx^idx^j,
\end{equation}
where $g_{ij}$ is the metric defined on $\so$ and the indices now have the range $1\le i,j\le n$. In both metrics $Q$ is globally hyperbolic provided $\so$ is complete, which is the case for the metric defined in \re{2.52}. The hyperbolic operator is symmetric in the first metric but not normally hyperbolic while it is normally hyperbolic but not symmetric in the second metric. Normally hyperbolic means that the main part of the operator is identical to the Laplacian of the spacetime metric.

Hence, if we want to describe quantum gravity not only by an equation but also by the metric of a spacetime then the metric in \re{6.18} has to be chosen. In this metric $Q$ has a curvature singularity in $t=0$, \cf \cite[Remark 6.3]{cg:uf2}. The Cauchy hypersurface $\so$ then corresponds to the hypersurface
\begin{equation}
\{t=1\}
\end{equation}
which also follows from the derivation of the quantum model where we consider a fiber bundle  $E$ with base space $\so$ and the elements of the fibers were Riemann metrics of the form
\begin{equation}
g_{ij}(t,x)=t^\frac4n\s_{ij}(x)
\end{equation}
where $\s_{ij}$ were  metrics defined in $\so$ and $t$ is the time coordinate that we use in $Q$, i.e.,
\begin{equation}
g_{ij}(1,x)=\s_{ij}(x).
\end{equation}
In the present situation we used the symbol $g_{ij}$ to denote the metric on $\so$ since $\s_{ij}$ is supposed to be a metric on $\Ss[2m-1]$.

Thus the event horizon is characterized by the Cauchy hypersurface
\begin{equation}
\{t=1\}.
\end{equation}
If $a=0$, i.e., in case we consider a Schwarzschild-AdS black hole as in \cite{cg:qbh}, then  we shall obviously assume that the black hole singularity
\begin{equation}
\{r=0\}
\end{equation}
corresponds to the curvature singularity
\begin{equation}
\{t=0\}
\end{equation}
of $Q$, i.e., the open black hole region is described in the quantum model by
\begin{equation}\lae{2.106}
\so\times (0,1)
\end{equation}
and the open exterior region by
\begin{equation}\lae{2.107}
\so\times (1,\un).
\end{equation}

If $a\not=0$, then there is  no curvature singularity in $r=0$, only a coordinate singularity in our present coordinate system. Indeed, if we choose generalized Boyer-Lindquist coordinates, \cf \cite[equation (3.1)]{gibbons:kerr-ads},  the metric has the form
\begin{equation}\lae{2.108}
\begin{aligned}
d\bar s^2&= -\frac{1+l^2r^2}{1-a^2l^2}d\tau^2+\frac{U r^2 dr^2}{(1+l^2r^2)(r^2+a^2) U -2 m_0 r^2}\\
&\q +\frac{r^2+a^2}{1-a^2l^2}\sum_{i=1}^m(d\m_i^2+\m_i^2(d\f_i+ l^2 d\tau)^2)\\
&\q+\frac{2m_0}U\big(d\tau-\frac a{1-a^2l^2}\sum_{i=1}^m\m_i^2d\f^i\big)^2.
\end{aligned}
\end{equation}
Then, defining
\begin{equation}
\rho= r^2,
\end{equation}
such that
\begin{equation}
d\rho=2 r dr
\end{equation}
we obtain new coordinates in which the metric is smooth up to $\rho=0$, indeed the metric is even smooth in the interval
\begin{equation}
-a^2<\rho<\un.
\end{equation}
In $\rho=-a^2$ there is curvature singularity:
\bl\lal{2.6}
The extended spacetime $N$ has a timelike curvature singularity in $\rho=-a^2$ .
\el
\bp
The fact that the curvature singularity is timelike follows immediately from \fre{2.49}, where we proved that outside the black hole region the hypersurfaces
\begin{equation}
\{\rho=\const\}
\end{equation}
are timelike.

To prove the existence of a curvature singularity  we first consider the case $m\ge 3$.  Looking at the metric in \re{2.108}  we observe that the components with respect to the coordinates $\mu_i$ form a diagonal matrix without any cross terms with the other coordinates, namely,
\begin{equation}\lae{2.113}
\begin{aligned}
\frac{\rho+a^2}{1-a^2l^2}\sum_{i=1}^md\m_i^2,
\end{aligned}
\end{equation}
where the $\m_i$ are subject to the side-condition
\begin{equation}
\sum_{i=1}^m\m_i^2=1,
\end{equation}
i.e., \re{2.113} represents the metric of a sphere of radius
\begin{equation}
\sqrt{\frac{\rho+a^2}{1-a^2l^2}}
\end{equation}
embedded in $\R[m]$ and the corresponding sectional curvatures in $N$ are defined independently of the other components of the metric in $N$ and they obviously become unbounded when $\rho$ tends to $-a^2$, since the sectional curvature $\s_p$ in a point $p\in N$ of a  plane spanned by two linearly independent vectors in
\begin{equation}
T_p(\Ss[m-1])\hra T_p(N)
\end{equation}
is equal to
\begin{equation}
\frac{1-a^2l^2}{\rho+a^2}.
\end{equation}
Secondly, in case $m=2$, we  used the package GREAT \cite{hubsch:great}  in Mathematica to compute the squared Riemannian curvature tensor in dimension $5$ and obtained
\begin{equation}
\bar R_{\al\bet\ga\de} \bar R^{\al\bet\ga\de}=\frac{96 m_0^2 \left(3 a^2-\rho\right) \left(a^2-3 \rho\right)}{\left(\rho+a^2\right)^6}+40 l^4
\end{equation} 
completing the proof of the lemma. 
\ep
 Since the curvature singularity is \tit{timelike} and not \tit{spacelike} as  the singularity of a Schwarzschild-AdS spacetime or the singularity in our quantum spacetime, equipped with the metric in \re{6.18},  it is easily avoidable. Despite this difference, we stipulate that the region in \re{2.107} corresponds to
\begin{equation}
\{\rho_2<\rho<\un\},
\end{equation}
where
\begin{equation}
r_2^2=\rho_2,
\end{equation}
and the region in \re{2.106} to
\begin{equation}
\{-a^2<\rho<\rho_2\}.
\end{equation}
\br
The time coordinate $\tau$ in a generalized Boyer-Lindquist coordinate system is a time function in
\begin{equation}\lae{3.72} 
N\sminus \bar B,
\end{equation}
where $N$ is the extended Kerr-AdS spacetime. We proved it directly  with the help of Mathematica,  if $\dim N=5$, by proving
\begin{equation}
\bar g^{\al\bet}\tau_\al\tau_\bet=\bar g^{00}<0.
\end{equation}
For a proof in any odd dimension  it will be sufficient to prove that the slices 
\begin{equation}
\{\tau=\const\}
\end{equation}
are spacelike in the region specified in \re{3.72}. Looking at the metric \re{2.108} we immediately see, by setting $d\tau=0$, that the induced metric is Riemannian.
\er
\br
When we have an extremal black hole with mass $m_0'$ and corresponding radius $r_0$ for the event horizon, then the function $\F=\F(\rho)$ in \re{2.12}, where
\begin{equation}
\rho=r^2,
\end{equation}
satisfies
\begin{equation}
0=\F(\rho_0)=\inf\F,
\end{equation}
hence 
\begin{equation}
\F'(\rho_0)=0.
\end{equation}
From the definition of $\F$ we then conclude that any black hole with mass
\begin{equation}
m_0>m_0',
\end{equation}
while the other parameters remain equal, will have an interior region. Hence, our previous arguments could then be applied to yield a quantum model depending on the Riemannian metric in \fre{2.50}. Letting $m_0$ tend to $m_0'$ the corresponding radii of the event horizons will then converge to $r_0$ leading to a quantum model for an extremal black hole.
\er

\br
In the quantum model of the black hole the event horizon is a regular Cauchy hypersurface and  can be crossed in both directions by causal curves hence no information paradox can occur. 
\er

%\backmatter
%\includepdf[pages=-]{/Users/claus/Documents/Scanned-Documents/}
\bibliographystyle{hamsplain}
%\bibliography{mrabbrev,publications}
\providecommand{\bysame}{\leavevmode\hbox to3em{\hrulefill}\thinspace}
\providecommand{\href}[2]{#2}

%\listoffigures

%\cleardoublepage

%\thispagestyle{empty}
%\closegraphsfile
\end{document}